\documentclass[prd,twocolumn,letterpaper,superscriptaddress,reprint]{revtex4}

\usepackage{graphicx}
\usepackage{pslatex}
\usepackage{amsmath}
\usepackage{amsbsy}

\newcommand{\beq}{\begin{equation}}
\newcommand{\eeq}{\end{equation}}

\newcommand{\atUCLA}{Dept. of Physics \& Astronomy, Univ. of California, Los Angeles, CA 90095.}
\newcommand{\atOSU}{Dept. of Physics, Ohio State Univ., Columbus, OH 43210.}
\newcommand{\atCCAP}{ Center for Cosmology \& Astroparticle Physics, Ohio State University.}
\newcommand{\atUH}{Dept. of Physics \& Astronomy, Univ. of Hawaii, Manoa, HI 96822.}
\newcommand{\atKU}{Dept. of Physics and Astronomy, Univ. of Kansas, Lawrence, KS 66045.}
\newcommand{\atWashU}{Dept. of Physics, Washington Univ. in St. Louis, MO 63130.}
\newcommand{\atNTU}{Dept. of Physics, Grad. Inst. of Astrophys.,\& Leung Center for Cosmology \& Particle Astrophysics, National Taiwan University, Taipei, Taiwan.}
\newcommand{\atSLAC}{SLAC National Accelerator Laboratory, Menlo Park, CA, 94025.}
\newcommand{\atUD}{Dept. of Physics, Univ. of Delaware, Newark, DE 19716.}

\newcommand{\atCFA}{Harvard-Smithsonian Center for Astrophysics,Cambridge, MA 02138 }
\newcommand{\atKICP}{Kavli Inst. for Comological Physics, Univ. of Chicago, Chicago, IL, USA, 60637}
\newcommand{\atChicago}{Dept. of Physics, Univ. of Chicago, Chicago, IL, USA, 60637}
\begin{document}

\title{Implications of ultra-high energy neutrino flux constraints\\ for Lorentz-invariance violating cosmogenic neutrinos}

\author{P.~W.~Gorham}
\affiliation{\atUH}

\author{A.~Connolly}
\affiliation{\atOSU}
\affiliation{\atCCAP}

\author{P.~Allison}
\affiliation{\atOSU}
\affiliation{\atCCAP}

\author{J.~J.~Beatty}
\affiliation{\atOSU}

\author{K.~Belov}
\affiliation{\atUCLA}

\author{D.~Z.~Besson}
\affiliation{\atKU}

\author{W.~R.~Binns}
\affiliation{\atWashU}


\author{P.~Chen}
\affiliation{\atNTU}
\affiliation{\atSLAC}

\author{J.~M.~Clem}
\affiliation{\atUD}





\author{S.~Hoover}
\affiliation{\atChicago}
\affiliation{\atKICP}

\author{M.~H.~Israel}
\affiliation{\atWashU}








\author{J.~Nam}
\affiliation{\atNTU}





\author{D.~Saltzberg}
\affiliation{\atUCLA}


\author{G.~S.~Varner}
\affiliation{\atUH}

\author{A.~G.~Vieregg}
\affiliation{\atCFA}



\begin{abstract}
We consider the implications of Lorentz-invariance violation (LIV) on cosmogenic neutrino observations,
with particular focus on the constraints imposed on several well-developed models for 
ultra-high energy cosmogenic neutrino production
by recent results from the Antarctic Impulsive Transient Antenna (ANITA) long-duration balloon
payload, and Radio Ice Cherenkov Experiment (RICE) at the South Pole. Under a scenario proposed originally by Coleman and Glashow, each lepton
family may attain maximum velocities that can exceed $c$, leading to energy-loss through
several interaction channels during propagation. We show that
future observations of cosmogenic neutrinos will provide by far the most
stringent limit on LIV in the neutrino sector. We derive the implied
level of LIV required to suppress observation of predicted fluxes from several mainstream cosmogenic 
neutrino models, and specifically those recently constrained by the ANITA and RICE experiments.
We simulate via detailed Monte Carlo code the
propagation of cosmogenic neutrino fluxes in the presence of LIV-induced energy losses.
We show that this process produces several detectable effects in the resulting attenuated neutrino spectra, 
even at LIV-induced neutrino superluminality of $(u_{\nu}-c)/c \simeq 10^{-26}$,
about 13 orders of magnitude below current bounds.
\end{abstract}
\maketitle


In the current Standard Model (SM) of particle physics, neutrinos are massless and unmixed, and couple to other
fundamental particles only through the weak interaction. This theory is very successful at
describing the gauge boson fields and their interactions, and the three generations of quarks and leptons.
However, observations of neutrino oscillations imply mixed neutrinos with non-zero mass, 
the most compelling evidence requiring extensions to the SM. 
The search for a fundamental theory that can explain and predict
properties of neutrinos with non-zero mass 
has in turn become one of the most active areas of high energy particle physics. It is thus natural to
anticipate that neutrino observations may provide unique opportunities for investigation of other beyond-standard-model
physics such as Lorentz invariance violation (LIV). While a recent claim of LIV in the form of
superluminal behavior in muon neutrinos~\cite{OPERA} appears to have been spurious and
due to subtle instrumental effects, the possibility that LIV can appear at very small levels 
in many sectors of particle physics has been seriously explored for decades. 
Investigation of LIV that could lead to non-tachyonic superluminal particle velocities 
has received renewed motivation from recent efforts to develop consistent quantum gravity theories,
through exploration of the possibility that Lorentz invariance may not be an exact vacuum symmetry at high energies~\cite{MT09,Horava09}.

SM modifications which specifically address LIV have been developed both as a way
of incorporating the low-energy effects of spontaneous CPT-violation into the SM~\cite{CK97},
and for high-energies, via general perturbation analysis~\cite{CG97,CG99}. The former work has been
generalized to a Standard Model Extension (SME), and as such it provides a 
comprehensive framework for analysis of almost any LIV effect in any particle
sector; recent compilations of experimental constraints now give limits for hundreds of
different possible LIV parameters~\cite{KM09,KR11}, and particular applications to the neutrino
sector, with a primary focus on low-energy phenomena, have also been developed
within this framework~\cite{KM04}, and more recently for even a fully relativistic 
formulation with operators of arbitrary dimension~\cite{KM12}.

The work of Coleman and Glashow~\cite{CG97,CG99}, 
which focused on the high-energy limit, provides some specific predictions regarding
energy-loss mechanisms for superluminal LIV charged particles. These results have been used
to provide stringent constraints on LIV on both ultra-high energy cosmic ray (UHECR) protons~\cite{CG97}
and UHECR photons~\cite{SG01}. More recently, similar considerations have been extended to 
cosmogenic neutrinos arising in intergalactic UHECR propagation
showing that LIV in the hadron sector could lead to suppression of photomeson production, and
a resulting decrease in the daughter neutrino fluxes~\cite{Scully10}. Thus detailed UHE cosmogenic neutrino measurements
and constraints can indirectly signal very small levels of LIV in the hadronic sector.


In the neutrino sector, LIV can appear in a very large variety of forms; for example, the recent 
comprehensive study by Kostelecky \& Mewes gives 369 possible coefficients of LIV parameters for models
which are restricted to be renormalizable~\cite{KM12}. In general, these coefficients can be associated with
both flavor-changing effects, where the velocity between neutrino mass eigenstates can differ, and
flavor-blind effects, where all flavors show an LIV-induced effect. In addition, directional dependence
of the particle velocity may be implied when the LIV involves a preferred inertial frame. 
The flavor-dependent
effects may appear as a type of oscillation signal for neutrinos propagating along different pathlengths or
along different directions, but need not imply superluminal LIV. Detection of an LIV-induced signal
due to these effects requires a search for modulation of the neutrino signal as a function of 
of the effective propagation distance, and such searches can be extremely sensitive since they
search for differential effects.
In contrast, flavor-blind effects would be responsible for producing LIV that would be unique to the entire neutrino
sector, affecting each mass eigenstate in the same way, and potentially leading to superluminal states. 
Searches for these effects must differentiate the neutrino sector velocity against 
another sector such as the photon sector, and are thus much more difficult in practice.

Currently,
velocity differences between different neutrino mass eigenstates are constrained to parts in $10^{-23}$ by
MINOS~\cite{MINOS10} with accelerator-based neutrinos, and to parts in $10^{-27}$ by IceCube using atmospheric muon neutrinos. 
However, constraints on the overall velocity difference of all neutrino flavors with
respect to photons are many orders of magnitude weaker, both than the constraints on intra-mass-eigenstate
variation, and compared to constraints on many other particle sectors in the SME. 

In this work we focus only on LIV
which can lead to an overall LIV superluminal behavior for the neutrino sector, and we will concentrate only on
isotropic effects without direction dependence. The Lagrangian density prescribed in the 
LIV-motivated SME leads to a modified energy-momentum relation
for neutrinos~\cite{CG99,Altschul09}:
\beq
E_{\nu}^2 =  m_{\nu}^2 ~c^4 + [1+2\delta_{\nu}(\hat{p}_{\nu})] ~c^2 ~\vec{p}_{\nu}^2
\eeq
where $E_{\nu}, m_{\nu}, \vec{p}_{\nu}$ are the particle's energy, mass, and momentum
and the parameter $\delta_{\nu}(\hat{p}_{\nu})$,
which is in general a spin-dependent, linear combination of several SME parameters,
and can be identified with the dimension-three isotropic coefficient 
in the Standard-Model Extension denoted $(c_{L})_{00}$, or $(c_{of})_{00}$ in the 
flavor-blind, oscillation-free limit~\cite{Diaz12,KM12}. 

This term effectively determines the maximum possible velocity in direction $\hat{p}_{\nu}$ as 
$u_{\nu} \approx c(1+\delta_{\nu})$,
where we have kept only terms to first order in $\delta_{\nu}$. 
This relation implies kinematically that the velocity of a neutrino 
may exceed that of photons {\em in vacuo} in the high-energy limit.
This behavior is distinct from the hypothesis of Lorentz-invariant {\em tachyonic} neutrinos,
first proposed by Chodos~{\it et al.}~\cite{Chodos85},
which have very different phenomenology and constraints; see~\cite{Davies12} for a recent review.

Recent work~\cite{CG11} has shown that LIV-induced superluminal neutrino propagation
in vacuum will lead to several energy-loss mechanisms not present in standard model
propagation.
One mechanism in particular, electron-positron pair creation, or {\it pair bremsstrahlung}, is 
efficient enough for even very small values of $\delta_{\nu}$, that it leads
to strong constraints in any experiment where neutrinos of sufficient energy
are detected after propagation over significant distances. This in turn implies
that even miniscule non-zero values of $\delta_{\nu}$ will effectively attenuate 
cosmogenic neutrino fluxes to undetectable levels. Detection of such effects 
take the form of a disappearance search, coupled with the possibility of
modification of the parent neutrino spectral energy distribution. This then alleviates
the need to compare the neutrinos against another particle sector to establish the
propagation characteristics.

Cowsik {\it et al.}~\cite{Cowsik12} have recently investigated the bounds on
LIV in the neutrino sector that arise from atmospheric muon neutrinos observed in
IceCube, using the  convention $E_{\nu} =  p_{\nu}  c(1+\alpha_{\nu})$  for the LIV parameter,
where they have neglected the neutrino mass in the high energy limit,
and thus $\alpha_{\nu} = \delta_{\nu}$ as used above in the SME. At high energies in the
pair bremsstrahlung process noted above, $\nu_i \rightarrow \nu_i' + e^- + e^+$, 
the neutrino loses about 3/4 of its energy, and thus to first order a single
pair bremsstrahlung interaction during neutrino propagation 
may be treated as an effective neutrino decay~\cite{CG11,Cowsik12}
with characteristic decay time
\beq
\tau_{\nu} =  \tau_{CG} ~E_{\nu,{\rm GeV}}^{-5}~ \alpha_{\nu}^{-3}~{\rm s}.
\label{tau}
\eeq
where $\tau_{CG} = 6.5 \times 10^{-11}$ s. For IceCube's observations of
upcoming atmospheric neutrinos, by requiring that the decay time exceed the
crossing time of the Earth, Cowsik {\it et al.} find a resulting bound
of $\alpha_{\nu} < 10^{-13}$.


Cowsik {\it et al.} also use similar analysis to derive a
first-order estimate of the possible upper limit to $\alpha_{\nu}$ that
would arise from any observation of UHE cosmogenic neutrinos, by again requiring
that the effective decay time be longer than the propagation time from the 
cosmogenic neutrino production site to Earth.
However, they considered only neutrinos arising from sources within the nearest 100~Mpc. 
While this is the most conservative approach, it is in tension with
the typical scenario presented by most UHE cosmogenic neutrino models.
In fact, in all current models for their production and propagation, 
only a negligible fraction of cosmogenic neutrinos observed at Earth
arise from nearby sources; by far the largest contribution is from relatively high-redshift sources.
To make an estimate consistent with this fact, we must first determine how to treat the
neutrino propagation and possible effective decay over cosmological distances.

\begin{figure}[htb!]
\includegraphics[width=3.5in]{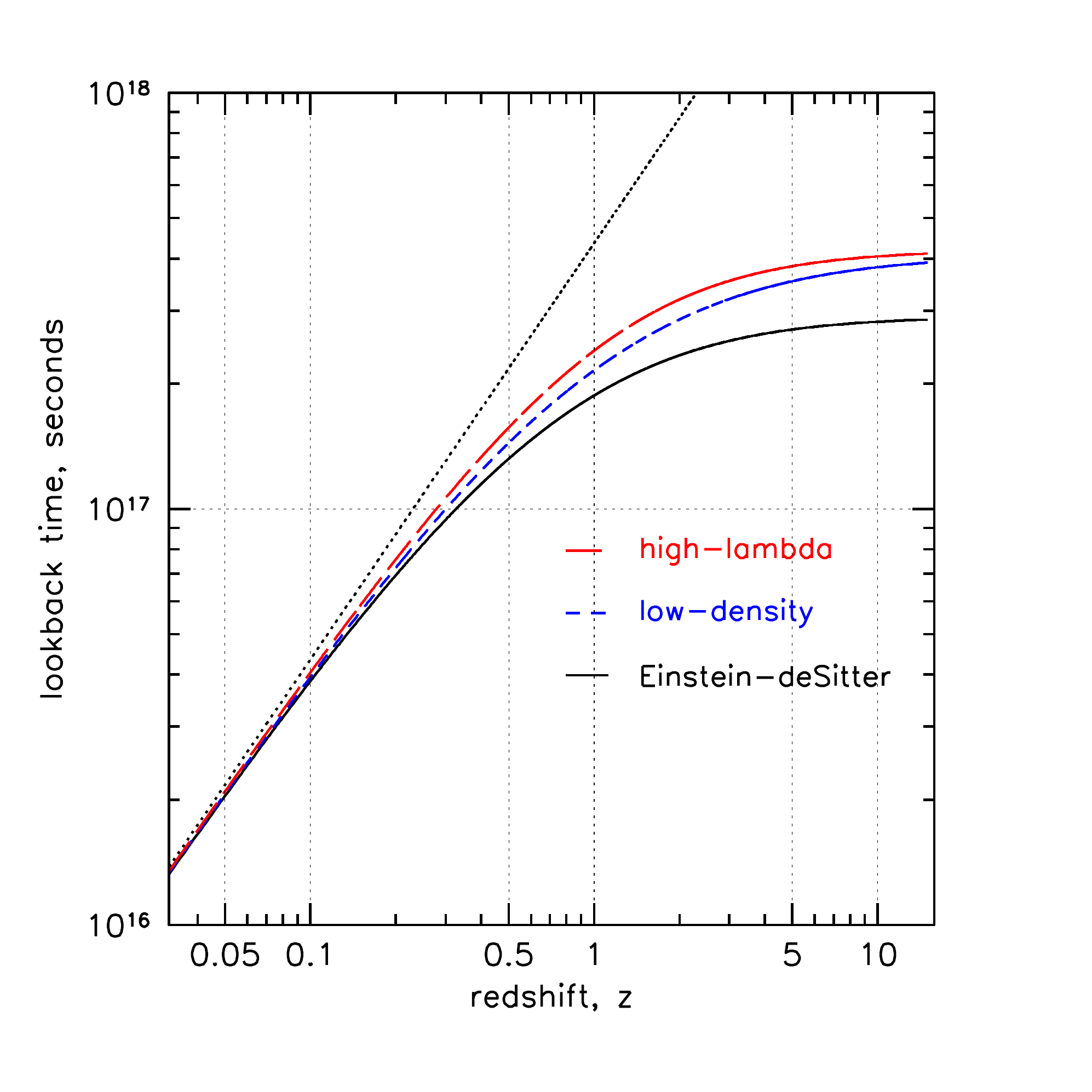}
\caption{Propagation time as a function of redshift $z$ for neutrinos for 3 different
cosmological models. The dotted line also indicates Euclidean propagation time.}.
\label{lbt}
\end{figure}

At a given redshift $z$, the propagation time for photons to the current
epoch is given by the {\it lookback time} $t_{\ell}$:
\beq
t_{\ell}(z) =  \int_0^z \eta(z')dz'
\eeq
where $\eta(z) = dt/dz = t_{H} [(1+z)\mathcal{E}(z')]^{-1}$~\cite{ESS}. Here
$t_{H} = H_0^{-1}$ is the Hubble time ($t_{H} \simeq 3.09 \times 10^{17}h$ sec
for $H_0 = 100h$ km s$^{-1}$ Mpc$^{-1}$, $h\sim 0.7$) and the function $\mathcal{E}(z)$ is
given by~\cite{Hogg}:
\beq
\mathcal{E}(z) = [\Omega_M (1+z)^3 + \Omega_k (1+z)^2 + \Omega_{\Lambda}]^{1/2}.
\eeq
Here the curvature parameter $\Omega_k = 1 - \Omega_M - \Omega_{\Lambda}$, where $\Omega_M,\Omega_{\Lambda}$
are the matter density and cosmological constant, respectively. We consider
three possible cosmologies, Einstein-deSitter ($\Omega_M = 1, \Omega_{\Lambda}=0$),
a low-density, high-curvature model ($\Omega_M = 0.05, \Omega_{\Lambda}=0$),
and the currently favored high-lambda model ($\Omega_M = 0.3, \Omega_{\Lambda}=0.7$).
Fig.~\ref{lbt} shows the lookback time for these three models, for $z>0.03$,
approximately 130~Mpc or more. The dotted line also shows the Euclidean-space approximation,
$t = z~t_H$. At $z>0.1$ departure from Euclidean propagation is evident, and at 
$z \sim 1$ the propagation time begins to saturate asymptotically
to the Hubble time, with less than a factor of two variation among the different 
cosmological models. 

A large number of studies of UHE cosmogenic neutrinos have been published since 1969
when Berezinsky and Zatsepin~\cite{BZ69} first described their origin as secondaries in
the Greisen-Zatsepin-Kuzmin (GZK~\cite{G,ZK}) process by which UHE cosmic ray protons
resonantly scatter off the cosmic microwave background photons. For our purposes here,
we refer only to several recent results that provide estimates of cosmogenic neutrino
spectra as observed at Earth, for a range of model parameters, and for sources
of different redshift~\cite{ESS,Kotera10}. 

Kotera {\it et al.}~\cite{Kotera10} present 
a decomposition of the neutrino energy spectral contributions from five different 
redshift bins, for $z < 0.5$,
$0.5 \leq z < 1.5$, $1.5 \leq z < 2.5$, $2.5 \leq z < 4$ and $z > 4$. This
decomposition of the spectral contributions is also given for six different 
cosmogenic neutrino models. In all cases, Kotera {\it et al.} find that 
the neutrino fluxes are dominated by contributions from $z>1.5$, well
into the asymptotic region of propagation time. Thus virtually all standard model cosmogenic neutrinos
have similar mean propagation times to Earth, $\langle t_{\nu} \rangle \sim 2-4 \times 10^{17}$ s, 
about a factor of 20-40 higher than that
assumed by Cowsik {\it et al.} in their estimate.

If a non-zero cosmogenic neutrino flux were observed, then
solving equation~\ref{tau} above, and requiring that the neutrino lifetime
should exceed the propagation time $t_{\nu}$ then gives an upper bound on $\alpha_{\nu}$:
\beq
\alpha_{\nu}(E_{\nu}) < 4.0 \times 10^{-4} ~E_{\nu,{\rm GeV}}^{-5/3}~~ t_{\nu}^{-1/3}~.
\eeq
Inserting the typical value $t_{\nu} = 3 \times 10^{17}$~seconds, and converting energy
to EeV units (1 EeV = $10^{9}$ GeV) which are more appropriate for the cosmogenic neutrino range
\beq
\alpha_{\nu}(E_{\nu}) < 6 \times 10^{-25} \left ( \frac{E_{\nu}}{1~\rm EeV} \right )^{-5/3}
\eeq
which, as Cowsik {et al} note, is many orders of magnitude below the current best bound
from IceCube observations of atmospheric neutrinos.

In fact this analysis is incomplete. Since the observed energy at Earth $E_{\nu,obs}$ is
redshifted relative to the energy at the source, in fact $E_{\nu,obs} = E_{\nu,src}/(1+z)$. Given that the
dominant redshift range for cosmogenic neutrinos is $1<z<4$, the bounds
on $\alpha_{\nu}$ are factors of 3-15 lower; we can approximate this as
\beq
\alpha_{\nu}(E_{\nu}) < 2 \times 10^{-25}  \left ( \frac{\langle z \rangle E_{\nu,obs}}{1~\rm EeV} \right )^{-5/3}
\label{lim1}
\eeq
where $\langle z \rangle$ is the weighted-mean $z$ of the cosmogenic neutrinos for a given source,
and $1<\langle z \rangle<4$ is the valid range. This somewhat overestimates the bound since the
propagation does not all occur at high-$z$; a more precise estimate requires integrating
the energy dependence as a function of $z$, as we will show below, but equation~\ref{lim1}
is a reasonable approximation.

\begin{figure}[htb!]
\includegraphics[width=3.5in]{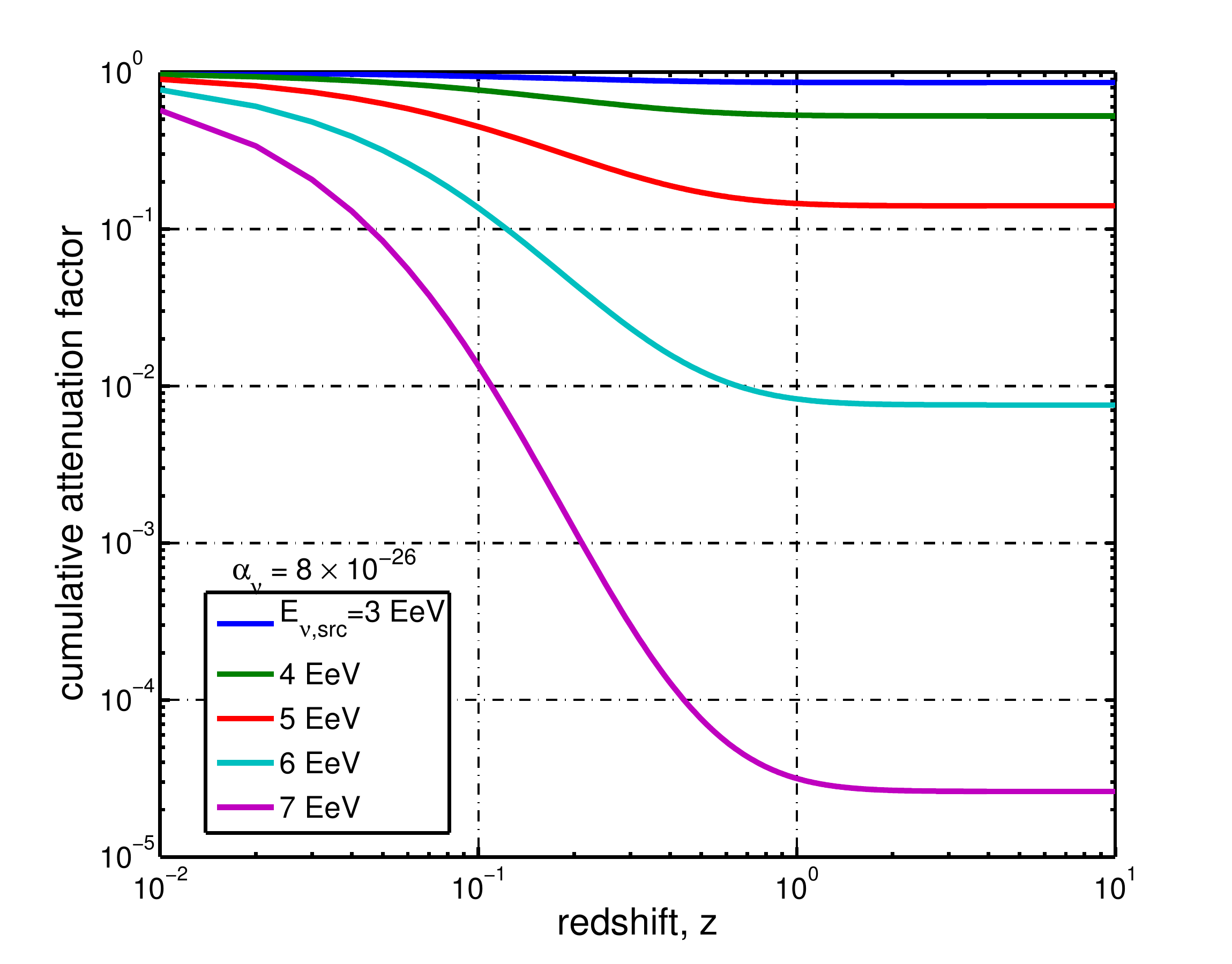}
\caption{The attenuation coefficient for a narrow range of source neutrino energies (eg. un-redshifted
energies) for the given value of $\alpha_{\nu}$, using the high-lambda cosmology and equation~\ref{Nz}.}.
\label{alphvE}
\end{figure}

Upper limits on fluxes of cosmogenic neutrinos do not of course uniquely point to 
new physics in order to suppress the flux, unless the limits begin to strongly
contradict other observations. For example, because the GZK process relies on
well-established particle physics with lab-frame energies in the GeV range, along
with the extremely well-measured cosmic microwave background radiation,
the observed UHECR fluxes guarantee an associated neutrino flux, if the primary
UHECRs are of a light composition, and the UHECR sources are distributed 
like other sources in the universe. A predominantly heavy (eg. iron) composition for the UHECRs,
such as that suggested by Auger Observatory measurements~\cite{AugerFe}, will
significantly suppress the cosmogenic neutrino fluxes, but only if the
the same UHECR composition measured for sources within the local universe applies
to cosmological sources at $z>1.5$. There is also currently 
some inconsistency in assuming a heavy composition for the primary UHECR, since
the observed GZK cutoff would require significant tuning of the source energy spectra
in order to match the cutoff, which would arise naturally for a light composition.


Despite these uncertainties, the infinitesimal level of LIV in the neutrino sector 
required to effectively kill the cosmogenic flux suggests that we should take this 
possibility seriously and consider its implications. Within the last several years,
a number of experiments, including the Radio Ice Cherenkov Experiment(RICE)~\cite{Rice},
IceCube~\cite{IceCubeHE}, and ANITA~\cite{ANITA-1,ANITA2}, have finally begun to constrain
fluxes of cosmogenic neutrinos. 
Although the total model space is 
not yet overly restricted by these constraints, the models that are ruled out,
termed {\em strong source evolutionary} models, are still consistent with
current UHECR data. These models do come into tension with indirect bounds derived 
by analyzing extragalactic gamma-ray data from the Fermi satellite~\cite{VB05,Gelmini12},
but these depend on details of the secondary cascade process and UHECR source
characteristics and these fluxes are thus not uniquely excluded~\cite{Scully10}. We 
therefore take these model bounds as motivation to investigate
what degree of LIV is required to
suppress these model fluxes below the current limits, and what other
observable implications this might have.

For $N$ neutrinos of energy $E_{\nu,src}$ propagating from the cosmogenic source to Earth, the
LIV-induced attenuation with propagation time is given by $dN/dt = -\lambda N$
with the decay length $\lambda = 1/\tau_{\nu}$. Then 
\beq
\frac{dN}{dz} = \frac{dN}{dt}\frac{dt}{dz} = - \frac{\eta(z)~ N}{~~\tau_{\nu}(E_{\nu}(z))}~.
\eeq
Substituting for $\tau_{\nu}$, rearranging and integrating both sides
\beq
N(z) = N_0~ \exp \left [ - \frac{\alpha_{\nu}^3}{\tau_{CG}}~ \int_0^z dz'~  \eta(z')~\left ( \frac{E_{\nu,src}}{1+z'} \right )^5 \right ]~.
\label{Nz}
\eeq
Figure~\ref{alphvE} shows an example of the attenuation coefficient that arises from this
propagation equation, for neutrino source energies over a quite small range, from 
3-7 EeV, and a value of $\alpha_{\nu}=8 \times 10^{-26}$ (chosen at random within the range of interest here) for the LIV parameter.
Here the redshift is that seen from the source looking out toward Earth, using the high-lambda model. It is evident that in each
case the attenuation has saturated by $z\sim 1$ relative to the source, which means that
any typical cosmogenic neutrino model will undergo the maximal attenuation of its beam toward Earth. 
We emphasize the very strong energy dependence displayed here, with attenuation for
a given value of $\alpha_{\nu}$ increasing by two orders of magnitude over an octave of energy.
This will tend to have a ``brick-wall'' impact on neutrino model fluxes, cutting them off
very rapidly above some onset energy, as we show below.

\begin{figure}[htb!]
\includegraphics[width=3.35in]{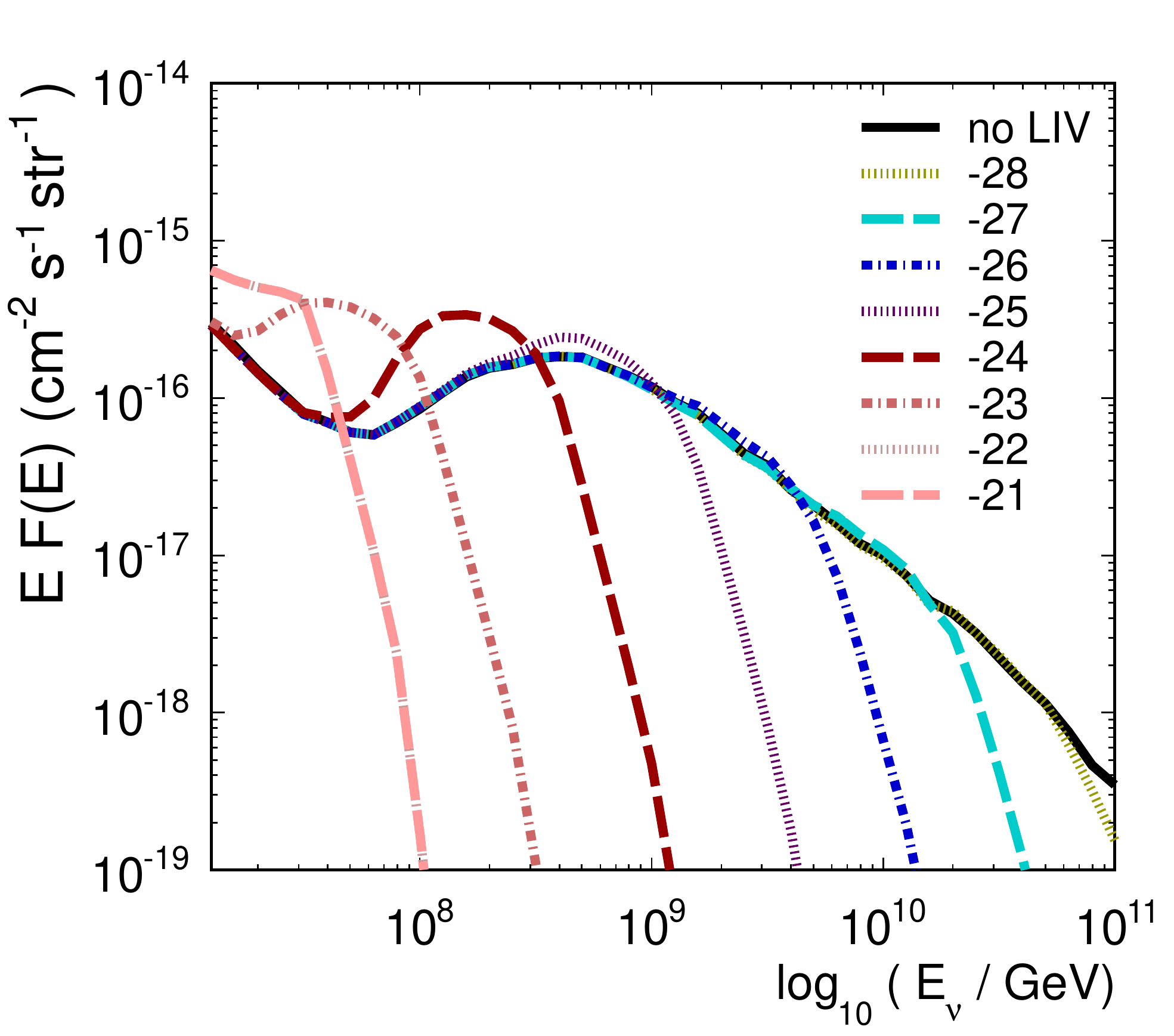}
\caption{Example of LIV effects on a strong-source-model cosmogenic neutrino spectrum. Black line:
unattenuated cosmogenic neutrino spectrum generated by our modified CRpropa code, 
here with parameters chosen to give a spectral shape commensurate with that of typical strong-source
cosmogenic neutrino models.
The different lines indicated in the legend show the value of $\log_{10}(\alpha_{\nu}$ that
yields the modified spectrum shown. In each case both the brick-wall cutoff at a given energy, and an enhancement
or pile-up effect just below that energy, appear in the LIV-modified spectra.
}
\label{amyLIV}
\end{figure}

We note also that the fact that there is a neutrino in the final state of this
interaction, carrying on average 1/4 of the energy, means that the spectrum will
not be simply attenuated above the cutoff energy, but that there should also be a pile-up
of lower energy neutrinos just at the edge of the effective LIV cutoff for a given
values of $\alpha_{\nu}$. Both the brick-wall and pile-up effects are
confirmed by an analysis from Bi {\it et al.} (2011)~\cite{XJBi2011}, where
much lower energy neutrinos are considered, over galactic propagation scales.



To determine the impact of both the energy-dependent attenuation and the 
potential final-state neutrino pileup on an UHE neutrino spectrum at earth, we use the 
publicly available cosmic-ray propagation code CRPropa~2.0~\cite{crpropa} to  simulate one-dimensional proton trajectories 
and then propagate the resulting neutrinos ourselves.
We simulate protons from cosmic monoenergetic point sources out to z=3~\cite{Conn12}.  
Using the energy and redshift of each neutrino produced along the trajectory, we allow 
it to interact via pair bremsstrahlung along its path to earth.  We subtract the cumulative 
attenuation factors determined by equation~\ref{Nz} from unity to obtain the energy-dependent probability density 
functions used to find the redshift of each neutrino interaction.
Multiple interactions are allowed from each initial neutrino.
Using the resulting collection of spectra from monoenergetic point sources, we follow the 
prescription for typical strong-source evolutionary models 
to calculate the neutrino spectra for an arbitrary injection spectrum and redshift evolution.  
We do not yet incorporate possible energy-dependence of the LIV parameter $\alpha_{\nu}$,
although such dependence may be expected from quantum gravity considerations~\cite{MT09};
in any case such effects would increase the sensitivity of our results to the
level of LIV.

\begin{figure}[htb!]
\includegraphics[width=3.2in]{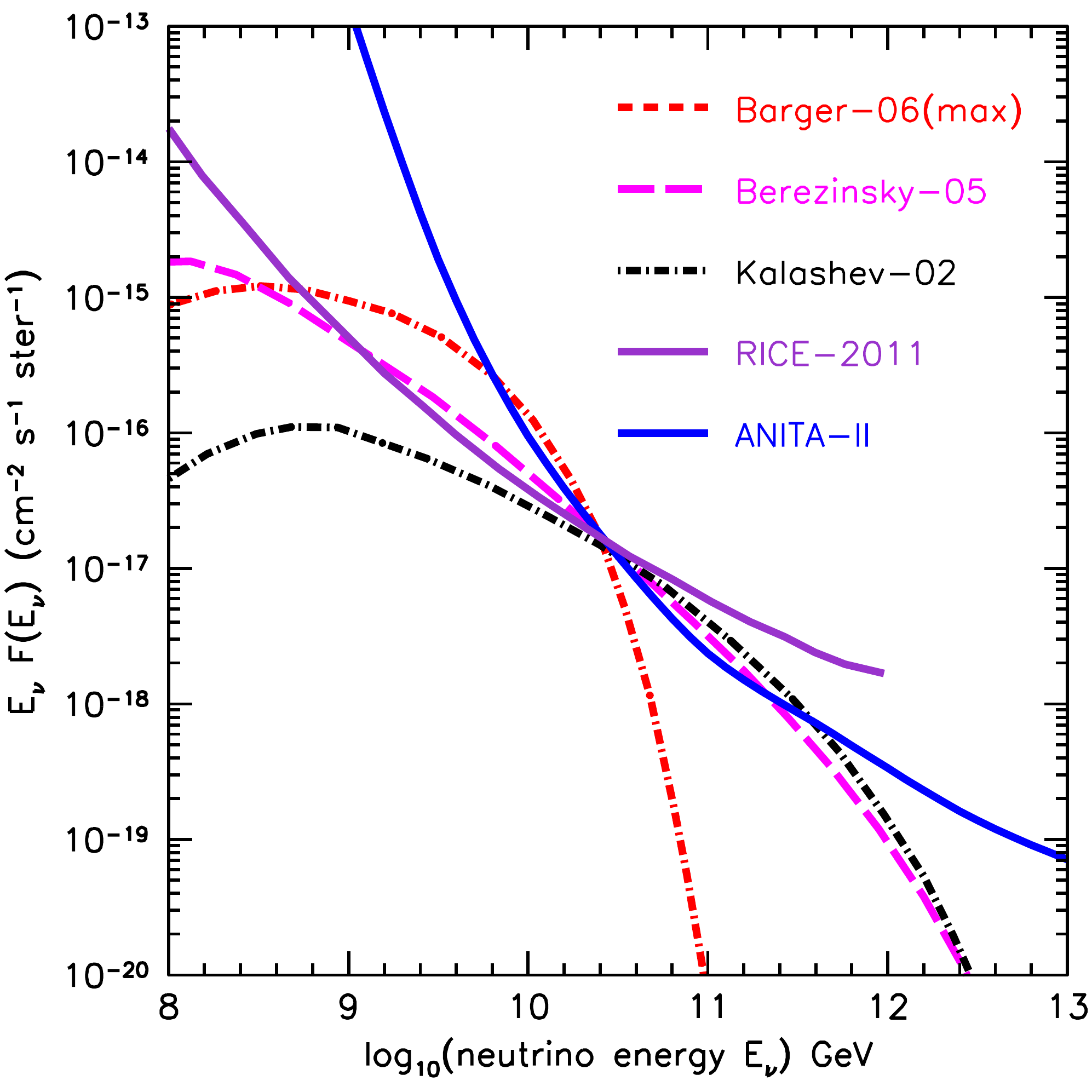}
\caption{ANITA-II and RICE limits and three strong-source-evolution model fluxes addressed here.}
\label{anitalim}
\end{figure}

For the final-state
outgoing neutrino, we assume the pair bremsstrahlung process as a mean 
inelasticity such that the final state neutrino's energy is 1/4 of the initial energy.
We would normally integrate over the inelasticity distribution, but such distributions are
not given in the current models. In general we have found that, at least for 
neutrino deep-inelastic scattering inelasticity distributions, using the mean value as
a proxy for a full integral is an acceptable approximation to first order.
Fig.~\ref{amyLIV} shows an example of these results, here applied to a 
cosmogenic neutrino model at varying levels of $\alpha_{\nu}$ ranging from
$10^{-27}$ to $10^{-21}$. Both the brick-wall and pile-up effects are evident,
and the turnover energy of the brick-wall in each case reflects the value of
$\alpha_{\nu}$ present.

Returning to the strong-source-evolution models currently constrained by ANITA \& RICE,
Fig.~\ref{anitalim} shows the fluxes for three of these models,
all of which are constrained above
the 90\% confidence level; also plotted are the ANITA and RICE limit curves. 
Energy flux (or intensity) units 
(particle energy $\times$ flux) are used here to avoid the spectral distortions that accompany 
less physically-motivated units such as $E^2 F(E)$ which are also often used in
reporting neutrino limits. In the normalization used here, 
models which just match the
differential limit curve for a decade of energy produce about 2 detectable events;
in these three cases, the expected number was between 4-6 events for ANITA for all three models. 
For RICE, we estimate that the highest model of Barger {\it et al.} would produce
of order 12 events; the higher-energy models will produce 2-3 events in RICE. 

\begin{table}[hbt!]
\caption{First column: expected numbers of events $N_{\nu}$ for ANITA-II \& RICE 
from three strong-source-evolution cosmogenic neutrino models; these models are all excluded at
$>90\%$ confidence from these experiments. Second column: values for the level of Lorentz invariance 
violation that would lead to the non-observation of these models.
The high-lambda cosmological model was employed for this calculation. 
\label{table1}}
\vspace{3mm}
 \begin{small}
  \begin{tabular}{lcc}
\hline \hline
{ {\bf Model \& references}}   &  ANITA/RICE~~&lower limit on \\ 
      &  predicted {$N_{\nu}$}~~& LIV $\alpha_{\nu}$ \\ \hline
Barger {\it et al.} 2006~\cite{Barger06} & &  \\
~~~~~ANITA-II & 3.5 &  $2 \times 10^{-27}$\\
~~~~~RICE2011 & 12 &  $2 \times 10^{-25}$\\
Berezinsky 2005~\cite{VB05} &   &   \\
~~~~~ANITA-II & 5.1 &  $3 \times 10^{-28}$\\
~~~~~RICE2011 & 3.4 &  $2 \times 10^{-27}$\\
Kalashev {\it et al.} 2002~\cite{Kal02} &  &   \\
~~~~~ANITA-II & 5.6 &  $2 \times 10^{-28}$\\
~~~~~RICE2011 & 2.9 &  $10^{-28}$\\
  \end{tabular}
 \end{small}
\end{table}

The event totals produced by an LIV-modified
spectrum are determined for ANITA using ANITA event-sensitivity
integrals based on our system Monte-carlos~\cite{ANITA-inst}; estimating these
event totals for RICE was outside the scope of our effort, but it is evident based on Fig.~\ref{anitalim}
that the constraints from RICE from the Barger {\it et al.} model could be a factor of
three better for that case at least. Table~\ref{table1} gives the hypothetical {\em lower limits}  on
$\alpha_{\nu}$ for each of the three cosmogenic neutrino models -- lower limits here
because they are the minimum values required to suppress these fluxes below the experimetal detection levels
for RICE and ANITA-II. Not surprisingly the model with the highest energy predictions 
has the highest sensitivity to the LIV parameter.  In all cases, even extremely
small values of LIV, far below existing limits in the flavor-blind case, would be adequate to make these cosmogenic
fluxes essentially undetectable, accounting for the non-observation of these fluxes.

It is an interesting coincidence that the level of LIV required to produce a flavor-blind superluminal
effect that would effectively suppress cosmogenic fluxes is comparable to the current best
limits on velocity differences {\em between} mass eigenstates, as noted in the introduction.
Of course there are other potential causes for suppression of the cosmogenic neutrino flux, and
thus detection of the cosmogenic neutrinos is clearly a more interesting outcome for both
neutrino astrophysics and for the resulting constraints on LIV phenomena.
We can thus estimate the strength of the upper bound on LIV for a hypothetical detection of
cosmogenic neutrinos for the sensitivity predicted
for ANITA-III, due to fly in 2013, under the assumption of near-zero background, as has been
the case for the first two ANITA flights. We have developed full Monte Carlo simulations of
the sensitivity of the ANITA-III payload currently in fabrication, and based on the planned
improvements we estimate about a factor of three increase in total neutrino event rate for a typical 30-day flight
compared to ANITA-II. Under this outcome, with a clear detection of several neutrino events, a
conservative constraint on
the LIV parameter from ANITA-III would be $\alpha_{\nu} \lesssim 10^{-26}$, which is consistent
with equation~\ref{lim1} above for $\langle z \rangle \sim 2$ and $E_{\nu,obs}\sim 3$~EeV.
Similar constraints would obtain for future detection of several unambiguous neutrino
candidates in RICE, or the future ARA experiment now in early construction~\cite{ARA},
or for the recently fully-instrumented IceCube experiment. Thus a detection of cosmogenic neutrinos
will move the flavor-blind constraints to values very close to those that currently obtain between the
mass eigenstates.


As noted above, a specific prediction of such LIV effects is the very steep dropoff with
energy once a threshold is reached. Thus, large-scale experiments such 
as IceCube~\cite{IceCubeHE}, Askaryan Radio Array (ARA)~\cite{ARA},
and others that have extended low-energy sensitivity to cosmogenic neutrino models as
compared to ANITA, may be able to detect such sharp cutoffs in neutrino energy spectra, signaling the onset of
LIV at these extremely small levels. Thus our results modify somewhat the conclusion of Cowsik {\it et al.}:
a clear detection of cosmogenic neutrinos {\em with no unexpected spectral features} would provide
exceedingly tight constraints on LIV in the neutrino sector. However, if enough neutrinos were
measured to indicate the very sharp energy-spectral cutoff predicted here, it would rather
unambiguously indicate the presence of neutrino LIV; no other proposed process
that we know of can produce such a sharp spectral cutoff in UHE neutrinos.
These results emphasize the need for a broad-spectral approach toward searches for the
cosmogenic neutrinos; this particular effect will turn on first at the highest energies,
and might not be apparent in a cosmogenic neutrino detector whose spectral reach did not extend beyond 10 EeV or more.

We conclude by observing that, although a relatively large value of $\alpha_{\nu}$ could completely
attenuate the cosmogenic neutrino fluxes below any detectable level for any experiment, the results of this
would not escape some other possible detection channels. Specifically, the pair bremsstrahlung
process will lead to electromagnetic cascades in intergalactic space, redistributing the
neutrino energy into an observable gamma-ray background. For example, in many models,
the photo-hadronic cascades that UHECRs undergo in the GZK process lead to rough equipartition
in energy between EeV cosmogenic neutrinos, and GeV cosmogenic gamma-rays. In the presence of
complete LIV-induced attenuation of the cosmogenic neutrinos, the cosmogenic gamma-ray
fluxes would be doubled to first order. Thus constraints on such
a process can be obtained through careful analysis of the extragalactic gamma-ray background,
such as that measured by Fermi, in direct analogy to what has already been done to
constrain the integrated intensity of the cosmogenic neutrino production~\cite{Gelmini12}. 
Such constraints may in fact be more robust than those on GZK-interaction cascades, since LIV-induced
neutrino cascades are much less influenced by uncertainties in magnetic fields
in the UHECR source environment.  

We are grateful to the US National Science Foundation Office of Polar Programs, the US Dept. of
Energy Office of Science, to NASA, and the Columbia Scientific Balloon Facility for their generous support of these efforts.
We thank the Ohio State University Supercomputing Center for the use of their facility to
generate the model neutrino spectra. We also thank Jorge Diaz for very useful comments regarding
the relation of our work to the Standard Model Extension.

\end{document}